\documentclass[sigconf,authordraf]{acmart}

\AtBeginDocument{
  }

\copyrightyear{2026}
\acmYear{2026}
\setcopyright{cc}
\setcctype{by}
\acmConference[ICSE '26]{2026 IEEE/ACM 48th International Conference on Software Engineering}{April 12--18, 2026}{Rio de Janeiro, Brazil}
\acmBooktitle{2026 IEEE/ACM 48th International Conference on Software Engineering (ICSE '26), April 12--18, 2026, Rio de Janeiro, Brazil}
\acmPrice{}
\acmDOI{10.1145/3744916.3773130}
\acmISBN{979-8-4007-2025-3/26/04}

\usepackage{xcolor}
\usepackage{soul}
\usepackage{enumitem}
\usepackage{float}
\usepackage{svg}
\usepackage{pifont}
\usepackage[utf8]{inputenc}
\usepackage{url}
\usepackage{fontawesome}
\usepackage{multirow}
\usepackage{colortbl}
\usepackage{listings}

\definecolor{Gray}{gray}{0.9}

\definecolor{light_red}{HTML}{fddad9}

\definecolor{javared}{rgb}{0.6,0,0}
\definecolor{javagreen}{rgb}{0.25,0.5,0.35}
\definecolor{javapurple}{rgb}{0.5,0,0.35}
\definecolor{javadocblue}{rgb}{0.25,0.35,0.75}

\lstdefinestyle{mystyle}{ 
    commentstyle=\color{codegreen},
    keywordstyle=\color{magenta},
    numberstyle=\tiny\color{codegray},
    stringstyle=\color{codepurple},
    basicstyle=\ttfamily\footnotesize,
    breakatwhitespace=false,         
    breaklines=true,                 
    captionpos=b,                    
    keepspaces=true,                 
    numbers=left,                    
    numbersep=5pt,                  
    showspaces=false,                
    showstringspaces=false,
    showtabs=false,                  
    tabsize=2,
    frame=none,
  xleftmargin=15pt,
  stepnumber=1,
  numbers=left,
  numbersep=5pt,
  stepnumber=1,
  numberstyle=\tiny\bf,
  belowcaptionskip=\bigskipamount,
  captionpos=b,
  escapeinside={*‘}{’*},
  tabsize=5,
  emphstyle={\bf},
  basicstyle=\scriptsize\ttfamily,
  keywordstyle=\color{javapurple}\bfseries,
  stringstyle=\color{javared},
  commentstyle=\color{javagreen},
  morecomment=[s][\color{javadocblue}]{/**}{*/},
  showspaces=false,
  columns=flexible,
  showstringspaces=false,
  morecomment=[l]{//},
  tabsize=2,
  breaklines=true，
}

\definecolor{lightyellow}{HTML}{FFFFCC}
\definecolor{lightblue}{HTML}{CCE5FF}
\definecolor{lightgreen}{HTML}{CCFFCC}

\definecolor{myhighlight}{HTML}{FCCD7D}

\newcommand{\headercolorlong}{\rowcolor{gray!17}}
\newcommand{\highlighttext}[1]{\colorbox{orange!30}{\textcolor{black}{#1}}}

\DeclareRobustCommand{\hlred}[1]{{\sethlcolor{light_red}\hl{#1}}}

\definecolor{Gray}{gray}{0.9}

\newcommand{\icse}[1]{#1}
\newcommand{\bonan}{}

\definecolor{light_green}{HTML}{c4edc9}
\definecolor{light_red}{HTML}{fddad9}

\DeclareRobustCommand{\hlred}[1]{{\sethlcolor{light_red}\hl{#1}}}
\DeclareRobustCommand{\hlgreen}[1]{{\sethlcolor{light_green}\hl{#1}}}

\definecolor{light_green}{HTML}{c4edc9}
\definecolor{light_red}{HTML}{fddad9}
\definecolor{light_gray}{HTML}{d5d5d5}

\DeclareRobustCommand{\hlred}[1]{{\sethlcolor{light_red}\hl{#1}}}
\DeclareRobustCommand{\hlgreen}[1]{{\sethlcolor{light_green}\hl{#1}}}

\let\oldfootnote\footnote
\renewcommand{\footnote}[1]{\oldfootnote{\textcolor{blue}{#1}}}
\newcommand{\dpr}{\underline{D}ense \underline{P}assage \underline{R}etrieval}
\newcommand{\tool}{\textsc{AutoDoc}}
\begin{document}

\pagenumbering{arabic}
\title{Automating API Documentation from Crowdsourced Knowledge}

\author{Bonan Kou}
\email{koub@purdue.edu}
\affiliation{%
  \institution{Purdue University}
  \city{West Lafayette}
  \state{Indiana}
  \country{USA}
}

\author{Zijie Zhou}
\email{zijiez4@illinois.edu}
\affiliation{%
  \institution{University of Illinois Urbana-Champaign}
  \city{Champaign}
  \state{Illinois}
  \country{USA}
}

\author{Muhao Chen}
\email{muhchen@ucdavis.edu}
\affiliation{%
  \institution{University of California, Davis}
  \city{Davis}
  \state{California}
  \country{USA}
}

\author{Tianyi Zhang}
\email{tianyi@purdue.edu}
\affiliation{%
  \institution{Purdue University}
  \city{West Lafayette}
  \state{Indiana}
  \country{USA}
}

\begin{abstract}
    API documentation is crucial for developers to learn and use APIs.
However, it is known that many official API documents are obsolete and incomplete.
To address this challenge, we propose a new approach called \tool{} that generates API documents with API knowledge extracted from online discussions on Stack Overflow (SO).
\tool{} leverages a fine-tuned dense retrieval model to \bonan{identify}  seven \bonan{types} of API knowledge from SO posts. Then, it uses GPT-4o to summarize the API knowledge in these posts into concise text. Meanwhile, we designed two specific components to handle LLM hallucination and redundancy in generated content.

We evaluated \tool{} against five comparison baselines on \bonan{48 APIs of different popularity levels}. 
Our results indicate that the API documents generated by \tool{} are up to {77.7\%} more accurate, {9.5\%} less duplicated, and contain {34.4\%} knowledge uncovered by the official documents. 
We also measured the sensitivity of \tool{} to the choice of different LLMs. 
\icse{We found that while larger LLMs produce higher-quality API documents, \tool{} enables smaller open-source models (e.g., Mistral-7B-v0.3) to achieve comparable results.}
\bonan{Finally, we conducted a user study to evaluate the usefulness of the API documents generated by \tool{}. All participants found API documents generated by \tool{} to be more comprehensive, concise, and helpful than the comparison baselines.} This highlights the feasibility of utilizing LLMs for API documentation with careful design to counter LLM hallucination and information redundancy.

\end{abstract}

\ccsdesc[300]{Information systems~Information extraction}
\keywords{knowledge extraction,software engineering,llm,api document}
\maketitle

\section{Introduction}
API documents are critical for software developers to write and debug code~\cite{liu2021api, treude2016augmenting, zhong2017empirical, uddin2019understanding}.
However, previous studies have shown that modern libraries and frameworks are plagued with incomplete and low-quality API documents~\cite{aghajani2019software, maalej2013patterns, robillard2011field, treude2020beyond, uddin2015api,huang2024planning}. 
Meanwhile, popular Q\&A websites such as Stack Overflow (SO) have accumulated a wealth of information on API usages~\cite{liu2021api}.
Compared with official API documentation, Q\&A posts are more up-to-date and comprehensive, covering various undocumented issues developers have encountered in practice. 

However, searching and sifting through SO posts for specific API knowledge is time-consuming, since such knowledge is often buried in many lengthy SO threads. 
To facilitate efficient knowledge retrieval from SO posts, several approaches have been proposed to extract API-related information from SO posts automatically~\cite{uddin2021automatic, treude2016augmenting, pandita2012inferring, chen2014asked}. Existing approaches are primarily heuristic-based or rely on supervised ML models. Such methods either lack the flexibility to extract knowledge from sophisticated natural language narratives or require a significant amount of labeled data. Curating such datasets requires significant human effort.

\begin{figure*}[!t]
    \centering
    \includegraphics[width=0.9\textwidth]{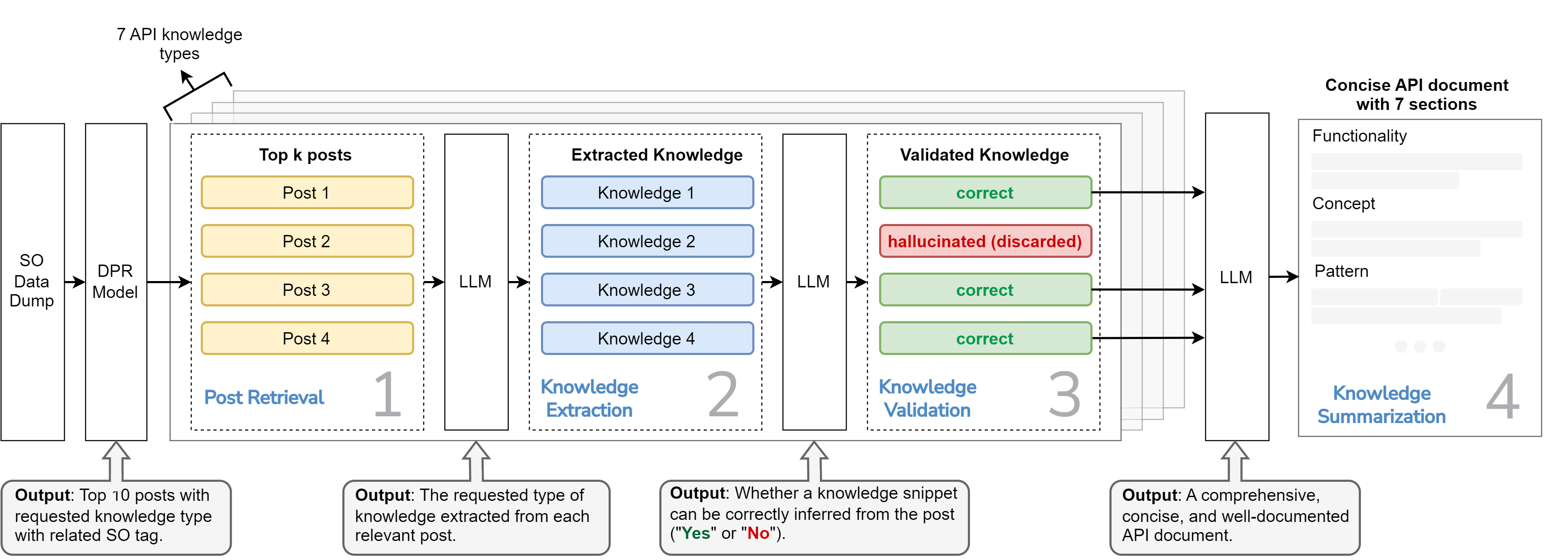}
    \caption{An Overview of the {\tool} pipeline.}
    \label{fig:pipeline}
\end{figure*}

In this work, we propose a new \bonan{approach} called \tool{} that leverages Large Language Models (LLMs) to extract API knowledge from SO posts. \tool{} generates well-organized API documents with seven types of API knowledge defined in the literature~\cite{maalej2013patterns, fucci2019using, liu2021api}.
Figure~\ref{fig:pipeline} shows the pipeline of \tool{}. 
First, given an API name and SO posts that are tagged with the API's library filtered from the entire SO data dump, \tool{} uses a {Dense Passage Retrieval} (DPR)~\cite{karpukhin2020dense} to identify relevant SO answer posts that may contain any of the seven types of API knowledge.
Since SO posts have unique linguistic patterns compared with general text, we fine-tuned the \bonan{original} DRP model with SO data.
Then, with the relevant posts retrieved, \tool{} uses GPT-4o to extract API knowledge from these posts. Since LLMs may generate misinformation (i.e., \textit{hallucination}), {\tool} performs {LLM-based self-checking} to infer the correctness of the extracted knowledge.
\bonan{Finally, \tool{}} prompts GPT-4o again to remove duplicates from the generated API documents. This step is necessary because the same knowledge can be extracted from multiple posts, \bonan{causing information redundancy.}

We evaluated \tool{} against five baselines, including two existing approaches in automated API documentation~\cite{8530028, treude2016augmenting}, and three prompting methods using GPT-4o~\cite{openai2023gpt4}.
We measure the percentages of knowledge snippets that are correct, unique, and do not overlap with the official API documents in the generated documents. We conducted experiments on a benchmark of 48 APIs with different popularity levels from Java, Android, Kotlin, and TensorFlow. 
The results show that API documents generated by \tool{} were up to {77.7\%} more correct and {9.5\%} more unique than API documents generated by the five baselines. Furthermore, on average, {34.4\%} knowledge snippets in API documents generated by \tool{} are not documented in the official API documents.

We also conducted an ablation study to measure the contribution of individual components of {\tool}. Our experiment results show that fine-tuning can improve DPR performance by {13\%}.
Without the knowledge validation component, the accuracy of API documents generated by \tool{} decreases by {3.7\%}. Without the knowledge summarization component, {31.7\%} more knowledge in documents generated by \tool{} is duplicated.  Furthermore, we replaced GPT-4o with three open source models (that is, Llama 3-70B-Instruct, Llama 2-70B-Chat, and Mistral-7B-v0.3) to assess the sensitivity of \tool{} to the choice of LLM. \icse{We found that while larger LLMs produce higher-quality API documents, \tool{} enables smaller open-source models to achieve comparable results. Specifically, the smallest LLM in our study, Mistral-7B-v0.3, can generate API documents that are 84.9\% accurate, 81.6\% unique, and contain 34.1\% knowledge snippets that are not covered by official API documents.}

Finally, we conducted a user study with {12} participants to evaluate the usefulness of the API documents generated by {\tool}.
Inspired by the {user study design from prior work}~\cite{7886920, xu2017answerbot,  nadi2020essential, 9401996}, we asked the participants to compare API documents generated by \tool{} and two baseline approaches in terms of \textit{helpfulness},
\textit{comprehensiveness}, \textit{conciseness}, \textit{overview}, and \textit{practice}. \bonan{In the study, 80\% participants preferred API documents generated by \tool{} in all five dimensions.}

In summary, we make the following contributions:

\begin{itemize}[left=0pt]
    \item We proposed a novel LLM-based pipeline for automated API documentation with a knowledge validation component that mitigates hallucination and a fine-tuned DPR model that effectively retrieves SO posts with API knowledge from SO data dump.
    \item We conducted a comprehensive evaluation, including a system-level comparison of \tool{} against five baselines, a detailed assessment of each component in {\tool}, and a user study with 12 participants.
    \item We have made the code and datasets publicly available to support reproduction and future research~\cite{icse2026_data}.
\end{itemize}


\section{Definition of API knowledge types}
\label{sec: formulation}
{\tool} aggregates different types of API knowledge shared on Stack Overflow into a comprehensive, well-organized document. Thus, before designing {\tool}, we first performed a literature review to identify the types of API knowledge that developers need when learning and using APIs~\cite{liu2021api, maalej2013patterns, huang2018api,thayer2021theory, meng2018application, dekel2009improving, ren2020demystify,li2018improving, monperrus2012should, zhang2019enriching}.  We summarize seven types of knowledge below.

\begin{itemize}[left=0pt]
\item \textit{Functionality} describes the actions or operations that an API can perform. 
Several studies have identified it as the most desired knowledge on Stack Overflow~\cite{liu2021api, maalej2013patterns, huang2018api}. 

\item \textit{Concept} refers to the abstract ideas and terminologies that an API aims to model. Previous studies have shown that knowing the concepts helps developers recognize relevant information and understand the purpose of the API~\cite{thayer2021theory, meng2018application, liu2021api, maalej2013patterns, huang2018api}.

\item \textit{Pattern} illustrates common use cases and code examples for using an API. 
Previous studies have shown that API usage patterns are essential for developers to learn new APIs~\cite{thayer2021theory, maalej2013patterns, liu2021api, zhang2019enriching}.
Specifically, in a user study with 54 participants, API usage patterns were found to significantly improve task completion rates in multiple API-related programming tasks~\cite{thayer2021theory}. 

\item \textit{Directive} provides guidelines on the proper use of an API, including best practices to follow and actions to avoid~\cite{ren2020demystify,maalej2013patterns,liu2021api, li2018improving, monperrus2012should}.
A previous study has identified directive knowledge as the third popular knowledge type in JDK documents~\cite{maalej2013patterns}.

\item \textit{Performance} refers to an API's time and memory efficiency. 
This type of knowledge is particularly useful in industrial settings with high standards for code performance~\cite{liu2021api, maalej2013patterns, dekel2009improving}.
For example, a previous study systematically surveyed several major APIs, including the JAVA standard library, Eclipse, and JMS, and found that performance knowledge for API usage is one of the most needed guidelines by developers~\cite{dekel2009improving}.

\item \textit{Environment} specifies the necessary conditions, system requirements, or configurations under which an API can function correctly~\cite{maalej2013patterns, robillard2011field, meng2018application}. While developers often need this type of knowledge to perform their work~\cite{de2004good, robillard2011field}, it is barely covered by official API documents~\cite{maalej2013patterns}.

\item \textit{Alternative} refers to other APIs with similar functionality, which can be considered {as} replacements or complementary options~\cite{liu2021api, dekel2009improving}.
For example, Liu et al.~found that 12.5\% of developers asked about alternative APIs on Stack Overflow~\cite{liu2021api}. 

\end{itemize}

\section{Approach}
This section describes the four key components of \tool{}: (1) \textit{relevant post retrieval},  (2) \textit{knowledge extraction}, (3) \textit{knowledge validation}, and (4) \textit{knowledge summarization}, as shown in Figure~\ref{fig:pipeline}.

\subsection{Relevant Post Retrieval}
\label{sec:data}

\begin{table*}[ht]
\small
\centering
\caption{Prompt to extract Functionality knowledge of Android \texttt{ActionBar} from Post 62606723 and the output from GPT-4o.}
\label{table:knowledge_extraction_example}
\begin{tabular}{>{\raggedright\arraybackslash\tt}p{0.98\textwidth}}
\toprule

\headercolorlong
\textbf{CONTEXT} \\
You are an API expert who excels at extracting, validating, and summarizing API knowledge from Stack Overflow posts as instructed. \\

\headercolorlong
\textbf{INSTRUCTION} \\
Extract functionality knowledge of ActionBar from this post, including knowledge you can infer from this post. If the post does not contain such knowledge, reply \textit{``No such knowledge''} and nothing more. \\

\headercolorlong
\textbf{API DESCRIPTION} \\
The ActionBar class in Android provides a dedicated space at the top of an activity for displaying the app's title, navigation controls, and other interactive elements like action buttons. \\

\headercolorlong
\textbf{KNOWLEDGE TYPE DESCRIPTION} \\
Functionality knowledge describes the actions or operations an API can perform. \\
\headercolorlong
\textbf{EXAMPLE KNOWLEDGE} \\
tf.gather is used to select tensor elements at specific indices.\\

\headercolorlong
\textbf{POST} \\
Use one of the ``.NoActionBar'' theme and set the windowActionBar attribute to false. Beginning with the API level 21 action bar may be represented with any toolbar widget, within the application layout. An application may choose to designate a Toolbar as the action bar for an Activity using the setSupportActionBar() method. \\

\hline
\headercolorlong
\textbf{GPT-4o Response} \\
1. Beginning with API level 21, the action bar can be represented by any Toolbar widget within the application's layout. \\
2. To designate a Toolbar as the action bar for an Activity, use the setSupportActionBar() method. \\

\bottomrule
\end{tabular}

\end{table*}

Since the SO data dump contains too much information to be passed into the LLM in one shot, 
\tool{} first leverages {\dpr{}} (DPR)~\cite{karpukhin2020dense} to identify SO answer posts that may contain each type of knowledge of a given API. 
DPR is a neural retriever that leverages contrastive learning to distinguish relevant and irrelevant information to a user query. 
Compared with other token-based or embedding-based retrieval methods, DPR has demonstrated superior performance and scalability to massive text corpora~\cite{karpukhin2020dense}. 

The original DPR model is trained on five popular open-domain QA datasets~\cite{karpukhin2020dense}.
Given the unique linguistic characteristics of SO posts, we fine-tuned the DPR model on a curated SO dataset to address the domain shift issue.
We constructed the fine-tuning dataset based on the insight that finding relevant posts with API knowledge is similar to finding answers to an SO question. 
Specifically, the dataset contains 144K SO question posts in the past five years and 175K corresponding answer posts with scores larger or equal to 5.
All data were filtered from the official Stack Exchange Data Dump~\cite{stackexchangearchive}.
During the fine-tuning process, we used each question title as a query. We used the answer posts to each question as positive passages and the answer posts to other questions as negative passages for contrastive learning. 
We used the Haystack framework for fine-tuning. We fine-tuned the DPR model for one epoch with a batch size of 16 and a gradient accumulation step of 8. The fine-tuning finished in 2 hours with 8 NVIDIA A5500 GPUs. 

For each API knowledge type of each API, \tool{} uses the DPR model to retrieve the top 10 answer posts that may contain such knowledge from the SO data dump.
We chose the number 10 to keep the API documents generated by \tool{} informative but not overwhelming. Queries used to retrieve posts that contain each type of knowledge can be found in our GitHub repository~\cite{icse2026_data}.

\subsection{Knowledge Extraction}
\label{sec:extraction}
Given the posts retrieved for each type of knowledge from the previous step, \tool{} uses GPT-4o to extract the corresponding knowledge from each post. 
We followed prior work~\cite{gao2024collabcoder, singh2023progprompt} to design a prompt template for few-shot prompting. Table~\ref{table:knowledge_extraction_example} illustrates the prompt used to extract the functionality knowledge of the \texttt{ActionBar} API in Android from a Stack Overflow post~\cite{Jabb2020} and the response from GPT-4o. The prompt includes six fields: (1) \textit{context information} that helps the model understand the domain we want it to operate within, (2) \textit{instruction} that asks the model to extract the required knowledge, {(3) \textit{API description} that describes the target API in one sentence}, (4) \textit{knowledge type description} that specifies which type of knowledge to extract, (5) \textit{knowledge example} that provides an example of the expected type of knowledge, (6) \textit{post} where knowledge should be extracted. 
These API descriptions were summarized by GPT-4o from the official API documents rather than manually written by the authors to avoid bias. The first author manually evaluated the correctness of these API descriptions and found no errors.
We present all API descriptions, API type descriptions, and knowledge examples on our GitHub repository~\cite{icse2026_data}.

Specifically, in the {\em instruction} field, the prompt asks the LLM to first determine whether the given post contains API knowledge or not before extraction. If the post does not contain the required API knowledge, the knowledge extraction model will output \textit{``No such knowledge''}. This design is inspired by the observation that the DPR model sometimes retrieves irrelevant SO posts. If the LLM responds \textit{``No such knowledge''}, \tool{} will consider this post irrelevant and discard it. Table~\ref{table:wrong_retrieval} shows an example of an incorrectly retrieved post for functionality knowledge of the \texttt{SAXParser} API. This post only mentions the API name but does not contain any actual functionality knowledge.

\begin{table}[htbp]
\caption{
A post that is incorrectly retrieved by the DPR model and rejected by the knowledge extraction component.}
\label{table:wrong_retrieval}
  \centering
  \normalsize
    \begin{tabular}{p{0.45\textwidth}}
      \hline
      \rowcolor{Gray}
      \textsf{ {Incorrectly retrieved post [Functionality, SAXParser]}} \\\hline 
      I have resolved this issue, it was due to another bug in my own code. So it had nothing to do with the SAXParser. Thanks for all the help!\\
      \hline
      \rowcolor{Gray}
      \textsf{GPT-4o response} \\ \hline
     \textit{No such knowledge.}\\
      \hline
    \end{tabular}
  
\end{table}


\subsection{Knowledge Validation}

\begin{table*}[ht]
\small
\centering
\caption{The prompt for validating a \textit{Performance} knowledge of Kotlin \texttt{ByteArray} from Post 65199270.}
\label{table:knowledge_validation_example}
\begin{tabular}{>{\raggedright\arraybackslash\tt}p{0.98\textwidth}}
\toprule

\headercolorlong
\textbf{CONTEXT} \\
You are an API expert who excels at extracting, validating, and summarizing API knowledge from Stack Overflow posts as instructed. \\

\headercolorlong
\textbf{INSTRUCTION} \\
Can this knowledge of ByteArray be extracted from this post? If so, reply "Yes". If not, reply "No". Just reply "Yes" or "No" and nothing more. \\

\headercolorlong
\textbf{API DESCRIPTION} \\
The ByteArray class in Kotlin represents an array of bytes, offering a fixed-size collection of byte values that can be accessed and manipulated by their indices. \\

\headercolorlong
\textbf{EXTRACTED KNOWLEDGE} \\
Reading all bytes of an InputStream into a ByteArray using the ``readBytes()'' extension method can be memory-inefficient for large inputs, as it loads the entire byte array into memory. \\

\headercolorlong
\textbf{POST} \\
The easiest way to make a ByteArray in Kotlin in my opinion is to use byteArrayOf(). It works for an empty ByteArray, as well as one which you already know the contents of. \\

\hline

\headercolorlong
\textbf{GPT-4o Response} \\
No. \\

\bottomrule
\end{tabular}
\end{table*}

Previous studies~\cite{rawte2023survey, zhang2023siren, huang2023survey} have shown that hallucination remains a major hurdle for adopting LLMs in downstream tasks. Since developers rely on API documentation to build software, it is critical to mitigate LLM hallucinations and ensure the reliability of generated API documents.
\icse{Recent work~\cite{ji2023towards} shows that LLMs can detect its own hallucination via self-reflection.} 
Inspired by this finding, {\tool} prompts the LLM to verify the correctness of the knowledge given in the post and discard the extracted knowledge that failed the verification.
Table~\ref{table:knowledge_validation_example} illustrates the prompt used to validate a performance knowledge snippet of \texttt{ByteArray} extracted from a SO post~\cite{Jabb2020} and the response from GPT-4o. The prompt includes five fields:
(1) \textit{context information} that helps the model understand the domain we want it to operate within, (2) \textit{instruction} that asks the model to validate the given knowledge based on the given post, (3) \textit{API description} that describes the target API in one sentence, (4) \textit{knowledge} that needs to be validated, and (5) \textit{post} based on which the knowledge should be validated.

\begin{table*}[ht]
\small
\centering
\caption{The prompt for summarizing knowledge of the \texttt{Model} API in Tensorflow.}
\label{table:knowledge_summarization_example}
\begin{tabular}{>{\raggedright\arraybackslash\tt}p{0.98\textwidth}}
\toprule

\headercolorlong
\textbf{CONTEXT} \\
You are an API expert who excels at extracting, validating, and summarizing API knowledge from Stack Overflow posts as instructed. \\

\headercolorlong
\textbf{INSTRUCTION} \\
Summarize these knowledge snippets regarding \texttt{Model} into a concise and accurate list. Focus on the seven types of knowledge as described below. For each type, create a section in the final API document. \\

\headercolorlong
\textbf{API DESCRIPTION} \\
A TensorFlow \texttt{Model} class defines a neural network's layers and forward pass for training and inference. \\

\headercolorlong
\textbf{ALL KNOWLEDGE TYPE DESCRIPTIONS} \\
KNOWLEDGE\_TYPE\_DESCRIPTIONS \\
\headercolorlong
\textbf{KNOWLEDGE LIST} \\
KNOWLEDGE\_TO\_BE\_SUMMARIZED \\

\bottomrule
\end{tabular}
\end{table*}

\subsection{Knowledge Summarization}
Since knowledge extracted from relevant posts may contain duplication, the knowledge summarization component is needed to eliminate redundant information. This component uses GPT-4o to summarize knowledge snippets extracted from multiple relevant SO posts into a concise API document.
This step is necessary because the relevant SO posts may contain duplicated information, and the same post can be retrieved for different knowledge types of the same API. Table~\ref{table:knowledge_summarization_example} illustrates the prompt used to summarize knowledge snippets of \texttt{Model} extracted from all relevant SO posts. The prompt includes five fields: (1) \textit{context information} that helps the model understand the domain we want it to operate within, (2) \textit{instruction} that asks the model to summarize the given knowledge snippets, (3) \textit{API description} that describes the target API in one sentence, (4) \textit{knowledge type descriptions} that explain each knowledge type, and (5) \textit{list of knowledge} that need to be summarized.

\section{Quantitative Evaluation}
We conduct experiments to answer seven research questions below:

\begin{itemize}[leftmargin=\parindent]
\item \textbf{RQ1}: How accurately and comprehensively can \tool{} generate API document \bonan{compared} to existing approaches, and how much knowledge can be found in the official documents?

\item \icse{\textbf{RQ2}: What leads to the incorrect information in generated documents?}

\item \icse{\textbf{RQ3}: To what extent do documents generated by different approaches complement each other?}

\item \textbf{RQ4}: What is the impact of API popularity on the effectiveness of \tool{}?

\item \textbf{RQ5}: To what extent does each component contribute to the effectiveness of \tool{}?

\item \textbf{RQ6}: How sensitive is \tool{} to the choice of LLM?

\item \textbf{RQ7}: What is the impact of temperature setting on the effectiveness of \tool{}?

\end{itemize}

\subsection{Comparison Baselines}
We compared \tool{} against five baselines, including API caveat~\cite{8530028}, SISE~\cite{treude2016augmenting}, and three prompting methods using GPT-4o~\cite{openai2023gpt4}. We elaborate on each of them below. 

\begin{itemize}[left=0pt]

    \item \textit{SISE}~\cite{treude2016augmenting} is an API document augmentation approach that selects insightful sentences with API knowledge from SO posts. It uses an SVM model that identifies insightful sentences based on both syntactic and semantic features. Since the source code and training data of SISE are not publicly available, we re-implemented SISE based on the approach descriptions in the paper. We curated a training dataset with 2,296 manually labeled sentences from SO posts related to the APIs used in our evaluation, as well as a test set with 460 labeled sentences. Our implementation of SISE achieves a precision of 62.5\%, which is comparable to the precision of the original SISE (60\%). Our implementation of SISE can be found in our GitHub repository~\cite{icse2026_data}.
    \item \textit{API caveat}~\cite{8530028} is a heuristic-based approach {for knowledge extraction from SO posts}. It uses 19 linguistic patterns to extract insightful sentences in SO posts. 
    The authors released the regex patterns used to capture these linguistic patterns, so we directly used these patterns in our paper.
    \item \textit{GPT-4o}~\cite{openai2023gpt4}. 
    Instead of using the sophisticated pipeline in \tool{} for API documentation, an alternative way is to directly prompt GPT-4o to generate an API document. We experimented with three prompting methods, including zero-shot prompting (ZSL), few-shot prompting (FSP), and chain-of-thought (CoT). Furthermore, to make it a fair comparison between GPT-4o and \tool{}, we asked GPT-4o to generate a document that includes the seven types of API knowledge defined in Section~\ref{sec: formulation} and included a description of each type of knowledge in the prompt. We present the prompts for these three GPT-4o baselines in our GitHub repository~\cite{icse2026_data}.
\end{itemize}

\subsection{Evaluation Benchmark}
We constructed an API benchmark with 48 APIs with different popularity levels and applied \tool{} and baselines to generate documents for these APIs. These APIs were selected from Java SDK, TensorFlow, Android SDK, and Kotlin SDK. These four libraries covered different topics in software engineering. For example, while Java and Kotlin are used for general purposes, Android SDK is used to develop mobile apps, and TensorFlow is used to develop machine learning projects. 
We performed stratified sampling to select APIs with different popularity levels.
API popularity was determined by its percentile ranking in GitHub code search. 
The popular APIs are from the top 10 percentile, the less popular ones are from 10 to 90 percentile, and the least popular ones are from the last 10 percentile. 
We randomly select 12 APIs for each of the 4 libraries. Among the 12 APIs, \icse{4} are high-popularity APIs, \icse{4} are medium-popularity APIs, \icse{4} are low-popularity APIs.
While 48 APIs may not sound like a large number,  it requires manual assessment of 48 APIs $\times$ 6 approaches = 288 API documents, which contains 3,840 knowledge snippets. Furthermore, the size of our benchmark is comparable to the evaluation benchmark of SISE~\cite{treude2016augmenting}, which includes 10 Java APIs. We present the complete list of the 48 APIs on our GitHub repository~\cite{icse2026_data}.





\subsection{\bonan{Evaluation Metrics and Data Labeling}}
\label{sec:metrics}
\icse{Inspired by existing research on API documentation~\cite{treude2016augmenting, treude2020beyond}, we use three metrics to measure the effectiveness of \tool{}. First, we follow SISE~\cite{treude2016augmenting} to manually label the \textbf{\textit{correctness}} of each knowledge snippet. Second, Treude et al.~\cite{treude2020beyond} find that a good API document should cover sufficient information briefly without redundant or unnecessary text. Thus, we measure the \textbf{\textit{uniqueness}} by counting the unique knowledge snippets in a document. Third, Boudin et al.~\cite{boudin2008scalable} show that it is important to provide information that developers cannot learn from the official documents. Thus, we measure the \textbf{\textit{overlap}} between an automatically generated document and the official document. All metrics reported in this paper are calculated as the total number of \textit{correct}/\textit{unique}/\textit{overlapping} knowledge snippets across all API documents in a given category (e.g., all documents generated by \tool{}), divided by the total number of knowledge snippets in those documents.}


The first two authors labeled the API documents generated by different approaches in two rounds.

\subsubsection{First round of labeling.} In the first labeling round, the two labelers independently identified and labeled all knowledge snippets in 5 API documents to familiarize themselves with the labeling tasks. These API documents were randomly sampled from 288 API documents generated by \tool{} and five baseline approaches. 
They started by highlighting all knowledge snippets in an API document. Specifically, each sentence in the document was labeled in four dimensions: (1) \textit{whether it is part of a knowledge snippet or not}, (2) \textit{whether it is correct or not}, (3) \textit{whether it is unique in the document or not}, and (4) \textit{whether it overlaps with the official API document or not}. We measured the agreement level of the two labelers using Cohen's Kappa. The Cohen's Kappa scores for the four dimensions are 0.60, 0.69, 0.84, and 0.58 respectively.  The labelers discussed their labels to establish the labeling guidelines outlined below:

For \textit{\textbf{Correctness}}, the labelers compared knowledge snippets in the generated API documents with online resources such as SO posts, blogs, tutorials, and official documentation. Partially correct snippets were considered incorrect. For example, a knowledge snippet that describes a usage of an API that will cause errors in specific versions of that API without mentioning this limitation will be labeled as incorrect.
Table~\ref{table:example_knowledge_snippets} shows an incorrect \textit{Pattern} knowledge of \texttt{ActionBar} and the correct version of the same knowledge snippet. 

\begin{table}[htbp]
  \caption{Different knowledge examples.}
  \label{table:example_knowledge_snippets}
  \centering
  \normalsize

    \begin{tabular}{p{0.45\textwidth}}
      \hline
      \rowcolor{Gray}
      \textsf{ \hlgreen{Correct} and \hlred{Incorrect} Knowledge Snippets} \\ \hline
      \hlgreen{1. Set titles in the ActionBar using \texttt{setTitle()} \textbf{[Correct]}} \\
      \hlred{2. Set titles in the ActionBar using \texttt{getSupportActionBar()} \textbf{[Incorrect]}} \\
      \hline
      \rowcolor{Gray}
      \textsf{Duplicated Knowledge Snippets} \\ \hline
      1. Later versions of \texttt{UUID} do not require system coordination. \\
      2. \texttt{UUID V2} does not require system coordination. \\
      \hline
      \rowcolor{Gray}
      \textsf{Overlapping Knowledge Snippet (Overlap \hlgreen{highlighted})} \\ \hline
      1. In that case, you should define your layers in \texttt{\_\_init\_\_()} and you should \hlgreen{implement the model's forward pass in  \texttt{call()}} when using \texttt{Model} \textbf{[Official documentation of \texttt{Model}]} \\
    2. \hlgreen{Use forward pass in \texttt{call()}} when using \texttt{Model} \textbf{[Knowledge extracted by \tool{}]} \\
      \hline
    \end{tabular}
  
\end{table}

For \textit{\textbf{Uniqueness}}, the labelers identified all duplicated knowledge snippets and reported the number of unique knowledge snippets in the document. A knowledge snippet was labeled to be duplicated if it completely paraphrased another snippet without providing any novel insight. Table~\ref{table:example_knowledge_snippets} shows two duplicated knowledge snippets where both imply \texttt{UUID V2} does not require system coordination. 

For \textit{\textbf{Overlap}}, a knowledge snippet from an auto-generated document was labeled as an overlap if it can be found in or inferred from the official API documents. Table~\ref{table:example_knowledge_snippets} shows an example of overlapping knowledge for the \texttt{Model} API.

\subsubsection{Second round of labeling} After they resolved the conflicts in their labels, the labelers started to label another five randomly sampled API documents following the established guidelines.
At the end of the second round, Cohen's Kappa scores for uniqueness, correctness, and overlap rose to 0.91, 0.95, and 0.88, indicating almost perfect agreement~\cite{mchugh2012interrater}. Therefore, the two labelers split and labeled the remaining 230 API documents separately. In total, the labeling process takes 102 person-hours.

\section{Quantitative Results}
\subsection{RQ1: Correctness and Comprehensiveness}
In Table~\ref{table:rq1}, Column \textsf{Correctness} shows the percentage of correct knowledge snippets in a generated document on average. Column \textsf{Uniqueness} shows the percentage of unique knowledge among the correct knowledge snippets in each document. Column \textsf{Overlap} shows the percentage of correct knowledge snippets that have been covered by the official API documents. Column \textsf{\# of snippets}  shows the total number of knowledge snippets in each document, regardless of correctness. 

In terms of \textit{\textbf{correctness}}, \tool{} generated the most accurate documents (96.2\%), outperforming traditional, non-LLM-based approaches (i.e., SISE and API caveat) by a large margin. The reason for the low performance of SISE and API caveat is that they are based on relatively simple rules or features and lack the capability to reason about the deep semantics of the natural language discussions in SO posts. Even though the SVM model used in SISE achieved 62.5\% accuracy in extracting insightful sentences on the original test set, its performance is not generalizable when applied to a broader scope of APIs and SO posts. This is a well-known generalizability issue of traditional, small models like SVM. 
    
It is surprising to see that simply prompting GPT-4o still produces API documents with decent accuracy, even with zero-shot prompting (90.4\%). It is known that GPT-4o is trained with a massive amount of Internet data, including official API documentation and Stack Overflow posts. Hence, this result implies that GPT-4o is very well-trained to memorize this data during pre-training. Yet supplementing external knowledge bases such as Stack Overflow posts still brings the benefit of generating more comprehensive API knowledge snippets. As shown in Column \textsf{Overlap} of Table~\ref{table:rq1}, when extracting knowledge from SO posts, \tool{} generates API documents with significantly less overlap compared to relying on GPT-4o alone. Furthermore, among the three prompting methods, chain-of-thought prompting achieves the highest correctness rate. This indicates that guiding the LLM to perform explicit reasoning in the generation process helps reduce hallucinations.

In terms of \textit{\textbf{uniqueness}}, \tool{} outperformed three prompting methods of GPT-4o by about 10\%. This is because \tool{} employs a knowledge summarization component that can effectively remove duplication in the document. 
Despite this, we noticed that API documents directly generated by GPT-4o contained more knowledge snippets compared to those generated by \tool{}. To investigate why, we manually examined the API documents generated by GPT-4o and found that GPT-4o tended to generate many code examples to illustrate how to use the API, which belongs to the type of {\em pattern} knowledge. Thus, they were labeled as valid knowledge snippets and contributed to the higher number of knowledge snippets in the documents generated by GPT-4o.

In terms of \textit{\textbf{overlap with official documents}}, \tool{} generates significantly fewer knowledge snippets that are covered by the official API documentation compared to the three prompting methods of GPT-4o. This is likely because GPT-4o relies on its internal knowledge to generate API documentation. Since GPT-4o is pre-trained on a massive amount of Internet data, including the official API documentation, GPT-4o has a tendency to spill out content it has seen during pre-training. By contrast, \tool{}, SISE, and API caveat rely on Stack Overflow as an external knowledge base and extracted API knowledge from SO posts, which are less likely to overlap with the official API documentation.

\begin{table}[]
    \caption{Comparison of \tool{} and baselines.}
\label{table:rq1}\resizebox{\linewidth}{!}{
\begin{tabular}{|l|rrrr|}
\hline
                                  & \multicolumn{1}{c}{\textbf{Correctness}} & \multicolumn{1}{c}{\textbf{Uniqueness}} & \multicolumn{1}{c}{\textbf{Overlap}} & \multicolumn{1}{c|}{\textbf{\# of Snippets}} \\ \hline
\textbf{SISE}                     & 18.6\%                                   & 90.1\%                                  & 56.3\%                               & 25.9                                          \\
\textbf{API caveat}               & 18.5\%                                   & 91.3\%                                  & 59.4\%                                 & 24.7                                          \\
\textbf{GPT-4o (ZSP)}             & 90.4\%                                   & 85.9\%                                  & 79.4\%                               & 32                                            \\
\textbf{GPT-4o (FSP)}             & 92.4\%                                   & 83.5\%                                  & 87.8\%                               & 32.8                                          \\
\textbf{GPT-4o (COT)}                    & 93.2\%                                   & 82.7\%                                  & 82\%                               & 33.8                                          \\
\textbf{\tool{}} & 96.2\%                                   & 93.2\%                                  & 65.6\%                               & 21                                            \\ \hline
\end{tabular}}
\end{table}

\subsection{RQ2: Manual Error Analysis}
\label{sec:error-analysis}

\icse{While \tool{} achieves 96.2\% accuracy, 38 knowledge snippets still include incorrect information. To understand what leads to such incorrect information, the first authors manually analyzed the 38 incorrect knowledge snippets and labeled the potential reasons. Each snippet was examined by checking the initially retrieved SO posts and the intermediate outputs generated by \tool{} at each step. Then, the first author discussed the analysis results with the other authors and categorized them into three types of errors.}

\icse{\subsubsection{\textit{SO post errors ({13.2\%})}} These errors stem from incorrect or misleading information in the retrieved SO posts. GPT-4o reuses this incorrect information from the original SO post when generating the new document. For example, the post in Table~\ref{table:post_error} suggests: \textit{``The point of UUIDs is that it doesn't matter who generates them...''}. This implies \texttt{UUID} does not require any system coordination, which is not true for \texttt{UUID} version 1 because it is generated using the MAC address and timestamp. GPT-4 reproduced this incorrect information when generating the document. 5 of the 38 incorrect knowledge snippets fall into this category.}

\begin{table}[htbp]
  \caption{Error due to inaccurate SO post}
  \label{table:post_error}
  \centering
  \normalsize

    \begin{tabular}{p{0.45\textwidth}}
      \hline
      \rowcolor{Gray}
      \textsf{An incorrect \textit{Pattern} knowledge snippet for \texttt{UUID}} \\ \hline
   Their generation does not require system coordination, enhancing
performance where distributed systems are involved.\\

      \hline
      \rowcolor{Gray}
      \textsf{Excerpt of SO post 73144845} \\ \hline
      The point of UUIDs is that it doesn't matter who generates them and there is no state you need to worry about.
      \\
      \hline
    \end{tabular}
  
\end{table}

\icse{\subsubsection{\textit{Misinterpretation of SO posts (36.8\%)}} These errors occur when GPT-4o misrepresents accurate content from SO posts—either by overgeneralizing, misclassifying, or extracting misleading summaries. For example, as shown in Table~\ref{table:gpt4o_error}, GPT-4o extracted a knowledge snippet claiming that \texttt{MediaPlayer} accepts input as long as it is encoded as an audio file. While this appears generally correct, it overgeneralizes the original intent of the SO post, which specifically discussed MP3 files. In fact, \texttt{MediaPlayer} only supports a limited set of audio formats recognized by Android (e.g., MP3 or MP4). {14} incorrect knowledge snippets fall into this category.}

\begin{table}[htbp]
  \caption{Error due to misinterpretation of SO post}
  \label{table:gpt4o_error}
  \centering
  \normalsize

    \begin{tabular}{p{0.45\textwidth}}
      \hline
      \rowcolor{Gray}
      \textsf{An incorrect \textit{Concept} knowledge snippet for \texttt{MediaPlayer}} \\ \hline
   \texttt{MediaPlayer} accepts input as long as it is still encoded as an audio file.\\

      \hline
      \rowcolor{Gray}
      \textsf{Excerpt of SO post 70556830} \\ \hline
      \textit{\textbf{Q:} Can I play MP3 files without a file extension in Android app?}\\
      \textbf{A:} \texttt{MediaPlayer} takes the input(the file) as long as it is still encoded as an audio file.
      \\
      \hline
    \end{tabular}
  
\end{table}

\icse{\subsubsection{\textit{Hallucinations without SO support ({50\%})}} These errors are fabricated information that cannot be traced back to any retrieved SO post or manually verified by any other information sources. Table~\ref{table:dpr_error} shows an example. GPT-4o hallucinated the existence of a \texttt{delete()} method in the \texttt{LinkedList} API. In reality, the standard Java \texttt{LinkedList} class does not define any method named \texttt{delete()}, and no retrieved SO post mentioned such functionality. {19} of the 38 incorrect knowledge snippets fall into this category.}

\begin{table}[htbp]
  \caption{Error due to hallucination without SO support}
  \label{table:dpr_error}
  \centering
  \normalsize

    \begin{tabular}{p{0.45\textwidth}}
      \hline
      \rowcolor{Gray}
      \textsf{An incorrect \textit{Functionality} knowledge snippet for \texttt{LinkedList}} \\ \hline
   \texttt{delete()} removes the last element from the linked list.\\

      \hline
      \rowcolor{Gray}
      \textsf{Excerpt of source post} \\ \hline
      No source post found.
      \\
      \hline
    \end{tabular}
  
\end{table}

\subsection{RQ3: Complementarity of Baselines}
\label{sec:overlap}

\label{sec:complementarity}

\icse{To systematically evaluate the complementarity of \tool{} with other baselines, we conduct a manual analysis on 10 randomly sampled APIs. For each API, the first author manually compared the knowledge snippets generated by \tool{} with all five baselines.}

\begin{table}[H]    
\caption{Knowledge from baselines uncovered by \tool{}.}
\label{table:complement}\resizebox{0.65\linewidth}{!}{
\begin{tabular}{|l|r|r|}
\hline
                      & \multicolumn{1}{c|}{\textbf{\# of Uncovered}} & \multicolumn{1}{c|}{\textbf{\# of Correct Snippets}} \\ \hline
\textbf{SISE}         & 2.7 (50.9\%)                                  & 5.3                                                  \\
\textbf{API caveat}   & 4.8 (60\%)                                    & 8                                                    \\
\textbf{GPT-4o (COT)} & 23.4 (82.9\%)                                 & 28.2                                                 \\
\textbf{GPT-4o (FSP)} & 21.8 (68.3\%)                                 & 30.9                                                 \\
\textbf{GPT-4o (ZSL)} & 25 (78.6\%)                                   & 31.8                                                 \\ \hline
\end{tabular}}
\end{table}

\icse{Table~\ref{table:complement} shows the number of knowledge snippets from baseline documents that are both correct and uncovered by \tool{}.
Compared with documents generated by GPT-4o, documents generated by API caveat and SISE contain fewer knowledge snippets that are not covered by documents generated by \tool{}. This is because they extract API knowledge from the same set of SO posts like \tool{}. Since GPT-4o is prompted to directly generate a document without referring to SO posts, GPT-4o mainly relies on its internal memory obtained from the pre-training data. As shown in Table~\ref{table:rq1}, about 80\% of the knowledge snippets in GPT-4o’s documents can be found in official documentation, whereas only 65.6\% of the snippets in \tool{}’s documents align with official sources. This difference explains why GPT-4o introduces more knowledge snippets that are absent from \tool{}’s results.}


\begin{table}[H]    
\caption{\tool{} on APIs of different popularities.}
\label{table:rq3}\resizebox{\linewidth}{!}{
\begin{tabular}{|l|rrrr|}
\hline
                & \multicolumn{1}{c}{\textbf{Correctness}} & \multicolumn{1}{c}{\textbf{Uniqueness}} & \multicolumn{1}{c}{\textbf{Overlap}} & \multicolumn{1}{c|}{\textbf{\# of Snippets}} \\ \hline
\textbf{High}   & 99.1\%                                   & 92.6\%                                  & 58.4\%                                 & 22.9                                           \\
\textbf{Medium} & 96.3\%                                   & 95\%                                  & 65\%                                 & 26.1                                         \\
\textbf{Low}    & 91.1\%                                   & 90.7\%                                  & 70.6\%                               & 14                                         \\ \hline
\end{tabular}}
\end{table}

\subsection{RQ4: APIs with Different Popularity Levels}
\label{sec:popularity}

Table~\ref{table:rq3} shows the performance of \tool{} for APIs with different popularity. In terms of \textbf{correctness}, we observed that the accuracy of knowledge snippets generated by \tool{} declines as the APIs become less popular. This is because less popular APIs are less frequently discussed on SO, providing fewer reference posts for \tool{} to extract API knowledge from when generating documentation. As a result, \tool{} relies more heavily on its pre-trained knowledge, which is prone to hallucination.
In terms of \textbf{uniqueness}, the performance of \tool{} is the worst for low-popularity APIs. This is because for APIs with low popularity, less relevant posts will be returned by the post retrieval component. In the case where one or more sections of API document corresponding to some knowledge types contain few or no knowledge snippets, the summarization component will fill these \icse{sections based on the model's internal memory. Table~\ref{table:dpr_error} shows an example where GPT-4o made up a knowledge snippet when the DPR model found no relevant posts.} 
In terms of \textbf{overlap}, we observe that for APIs with low popularity, \tool{} generated more knowledge snippets that can be found in official API documents. A possible reason is that less popular APIs were less frequently discussed on SO. Therefore, it is hard for the DPR model to retrieve posts that contain undocumented knowledge of these APIs. Without enough reference posts, LLMs like GPT-4o could elaborate on some knowledge snippets when asked to summarize them. In that case, the extra content elaborated by the LLM (e.g., explaining a parameter used in an API) can only come from the LLM's pre-trained knowledge, which can overlap with official API documents.

\subsection{RQ5: Contribution of Each Component}
In this section, we report the contribution of each component of \tool{}: (1) fine-tuned DPR model, (2) knowledge extraction component, (3) knowledge validation component, and (4) knowledge summarization component.

\subsubsection{Fine-tuned DPR model}
\label{sec:dpr_evaluation}
To evaluate the performance of the fine-tuned DPR model, we compared it against the original DPR model and two popular retrieval methods, BM25~\cite{robertson2009probabilistic} and E5 (an embedding model)~\cite{wang2022text}. The evaluation was performed on 10 APIs randomly sampled from our benchmark. For the retrieved posts for these 10 APIs, please check our GitHub repository~\cite{icse2026_data}.

The first two authors who labeled the generated API documents also labeled the retrieved posts. Specifically, they labeled 10 APIs × 7 knowledge types × 10 retrieved posts per type × 4 retrievers = 2,800 retrieved posts. They first independently labeled 100 posts randomly selected from posts retrieved by both models. Each post can either be relevant or irrelevant to the DPR query. In the first labeling round, the two labelers achieved a Cohen's Kappa score of 0.63. They discussed their labels to establish a labeling rule that a post should be labeled as relevant only if it contains the requested type of knowledge. With that rule in mind, a post containing only \textit{functionality} knowledge of an API should be labeled as irrelevant if it was retrieved by the query of \textit{concept} knowledge. After resolving all conflicts in their labels, the two labelers independently labeled another 100 randomly sampled posts, achieving a Cohen's Kappa score of 0.74. Since this score indicated a substantial level of agreement, they split and labeled the remaining posts separately. In total, the labeling process took 17.2 person-hours.
\begin{table}[H]    
\caption{Performance of different retrieval methods.}
\label{table:dpr_result}
\resizebox{0.65\linewidth}{!}{
\begin{tabular}{|l|r|r|r|r|}
\hline
\textbf{Knowledge} 
& \multicolumn{1}{c|}{\textbf{E5}} 
& \multicolumn{1}{c|}{\textbf{BM25}} 
& \multicolumn{1}{c|}{\textbf{DPR}} 
& \multicolumn{1}{c|}{\textbf{DPR (fine-tuned)}} 
\\ \hline
Functionality & 7\% & 35\% & 26.9\% & 41\% \\
Concept       & 13\% & 47\% & 36\% & 61\% \\
Performance   & 2\% & 45\% & 33\% & 42\% \\
Directive     & 9\% & 34\% & 33\% & 38\% \\
Pattern       & 16\% & 50\% & 45\% & 56\% \\
Environment   & 9\% & 46\% & 36\% & 48\% \\
Alternative   & 10\% & 35\% & 38\% & 53\% \\
\hline
\textbf{Overall} & 9.4\% & 41.7\% & 35.4\% & 48.4\% \\
\hline
\end{tabular}}
\end{table}

Table~\ref{table:dpr_result} shows the average accuracy of the fine-tuned DPR model and three baselines on retrieving the top 10 most relevant posts for the 10 APIs randomly selected from our benchmark by knowledge type.
Overall, the fine-tuned DPR model outperforms all baselines by a large margin. Surprisingly, BM25 outperforms the original DPR model with no fine-tuning. This is because the original DPR model does not understand the technical language in SO posts. On the other hand, BM25 uses term frequency-based scoring, allowing it to effectively capture relevant API names.  
We notice that fine-tuning increased the retrieval accuracy of the DPR model on all knowledge types by 13\%. This indicates fine-tuning can effectively improve the performance of the DPR model on domain-specific tasks.
Specifically, the retrieval accuracy for \textit{concept} (+15\%) and \textit{alternative} knowledge (+15\%) benefited the most from the fine-tuning. A possible explanation is that retrieving \textit{concept} and \textit{alternative} knowledge involves recognizing how different API components interact, their intended use cases, and possible substitutions, which is too hard for the original DPR without finetuning.

\subsubsection{Knowledge extraction component} 
To evaluate the accuracy of the knowledge extracted by \tool{} from SO answer posts, we randomly sampled 321 knowledge snippets from 1,920 knowledge extracted from 3,360 SO answer posts (10 posts retrieved for each knowledge type/API combination). This sample size is statistically meaningful with a 95\% confidence interval. For each API knowledge, the first two authors independently verified whether the knowledge was accurate by searching online. Then, they compared their verification results and resolved any disagreement. The Cohen's Kappa score was 0.84. Overall,  301 of the 321 relations were verified to be correct, indicating a high accuracy (93.7\%).

\begin{table}[htbp]
\caption{
An identified incorrect \textit{Alternative} knowledge for \texttt{MediaPlayer} from Post 76461455.}
\label{table:validation_example}
  \centering
  \normalsize

    \begin{tabular}{p{0.45\textwidth}}
      \hline
      \rowcolor{Gray}
      \textsf{ {POST 76461455}} \\ \hline
      On \highlighttext{Android}, you can use the MediaPlayer class to play sound. The MediaPlayer class can play audio and video files, and it provides a rich set of features such as seeking, looping, and volume control. \\
On \highlighttext{iOS}, you can use the AVFoundation framework to play sound. The AVFoundation framework provides classes for playing, recording, and editing audio and video. To play a sound on iOS, you can use the AVPlayer class. \\
      \hline
      \rowcolor{Gray}
      \textsf{EXTRACTED ALTERNATIVE KNOWLEDGE} \\ \hline
      Alternatives to MediaPlayer on Android include the AVPlayer class from the AVFoundation framework on \highlighttext{iOS} for playing audio and video. \\
      \hline
      \rowcolor{Gray}
      \textsf{VALIDATION RESULT} \\ \hline
      No.\\
      \hline
    \end{tabular}
  
\end{table}

\subsubsection{Knowledge validation component}
We conducted an ablation experiment where the knowledge validation component was removed from \tool{}.
This ablation experiment was conducted on 10 APIs randomly sampled from our benchmark. Table~\ref{table:effect_of_validation_and_summarization} shows the results of this ablation study. Our experiment showed that the accuracy of knowledge snippets in the generated API document, on average, dropped 3.7\% after removing it. This result indicated that the knowledge validation component can effectively remove hallucinations from the generated API documents. 
For example, Table~\ref{table:validation_example} shows an example where the knowledge validation component identified an incorrect alternative knowledge of \texttt{MediaPlayer} in Android extracted by GPT-4-o from Post 76461455. Although the post mentioned \texttt{AVPlayer} can be used to play sound as well, it cannot replace \texttt{MediaPlayer} because they were used on different operating systems (Android v.s.~IOS). Therefore, as the knowledge validation component correctly concluded,  Post 76461455 did not contain alternative knowledge of \texttt{MediaPlayer}.

Furthermore, we noticed that when the validation component was removed, the generated API document also became more duplicated. For example, the API documents became 15.2\% more duplicated on average. This is because the knowledge validation component helps to remove hallucinated content extracted from a relevant post. Most of the hallucinated content filtered by the validation component was not factually wrong but could not be found in the relevant post and, therefore, should not be extracted. Such hallucinated contents are usually simple, general knowledge of an API (i.e., \textit{``Model in TensorFlow is used to define and create neural network models''}), which can easily overlap, introducing duplication.

\subsubsection{Knowledge summarization component}

We also conducted an ablation study where the knowledge summarization component was removed from the pipeline. This ablation study was conducted on the same 10 APIs used to evaluate the knowledge validation component. Our experiment results (Table~\ref{table:effect_of_validation_and_summarization}) showed that the generated API documents, on average, became 31.7\% more duplicated after removal.
This indicates the effectiveness of the knowledge summarization component. 
For example, Table~\ref{table:summarization_example} shows three duplicated knowledge snippets extracted from different posts and the summary made by the summarization component. The same concept knowledge, \textit{``ByteArray is an array of bytes in Kotlin''}, was repeatedly extracted from different posts, causing redundancy in the generated document. The knowledge summarization component successfully detected and removed this duplication.

\begin{table}[H]
\captionsetup{aboveskip=0pt, belowskip=0pt}
\caption{Effect of knowledge validation and summarization.}
\label{table:effect_of_validation_and_summarization}
\resizebox{\linewidth}{!}{
\begin{tabular}{|l|rrrr|}
\hline
                                  & \multicolumn{1}{c}{\textbf{Correctness}} & \multicolumn{1}{c}{\textbf{Uniqueness}} & \multicolumn{1}{c}{\textbf{Overlap}} & \multicolumn{1}{c|}{\textbf{\# of Snippets}} \\ \hline
\textbf{\tool{}} & 96\%                                     & 87.6\%                                  & 66.7\%                               & 17.2                                         \\
\textbf{- validation}             & 92.3\%                                   & 75.8\%                                  & 72\%                               & 18.2                                         \\
\textbf{- summarization}          & 94.1\%                                   & 59.3\%                                  & 74.5\%                               & 21.7                                         \\
 \hline
\end{tabular}}
\end{table}

\vspace{-7pt}
\begin{table}[htbp]
\caption{
A list of duplicated knowledge snippets of \texttt{ByteArray} and the summarized version.}
\label{table:summarization_example}
  \centering
  \normalsize

    \begin{tabular}{p{0.45\textwidth}}
      \hline
      \rowcolor{Gray}
      \textsf{ {EXTRACTED KNOWLEDGE SNIPPETS}} \\\hline 1. ByteArray is an array of bytes in Kotlin.\\
2. A ByteArray in Kotlin is an array of bytes.\\
3. ByteArray is a data structure in Kotlin that represents a sequence of bytes.\\
      \hline
      \rowcolor{Gray}
      \textsf{SUMMARIZATION RESULT} \\ \hline
     ByteArray is used to create and manage arrays of bytes in Kotlin.\\
      \hline
    \end{tabular}
  
\end{table}

\subsection{RQ6: Sensitivity to Different LLMs}
In addition to GPT-4o, we also experimented with using other LLMs as the underlying LLM for \tool{}. We selected three open-sourced LLMs, including Llama 2-70B-Chat, Llama 3-70B-Instruct, and Mistral-7B-v0.3. We evaluated the performance of these three variants of \tool{} on 10 APIs randomly sampled 
 from our benchmark and showed the results in Table~\ref{table:llms}. After shifting from GPT-4o to open-sourced LLMs, the percentage of correct knowledge snippets and unique knowledge snippets both dropped to different extents. This is expected because GPT-4o is one of the state-of-the-art LLMs. Among the three open-sourced LLMs, Llama 3-70B-Instruct generated the most correct and unique API documents. 
This is because Llama 3-70B-Instruct was pre-trained on 8.3 times more tokens than Llama 2-70B-Chat~\cite{grattafiori2024llama} and has way more parameters than Mistral-7B-v0.3. 
Surprisingly, despite being the smallest model, Mistral-7B-v0.3 achieved similar performance with Llama 2-70B-Chat and outputted the most number of knowledge snippets (21.8). This is likely due to the Mixture of Experts (MoE) architecture~\cite{jacobs1991adaptive} in Mistral-7B-v0.3, which compensates for its small size to some extent. Also, API documents with more knowledge snippets are not necessarily better. LLMs can generate short bullet points, each of which will be labeled as a knowledge snippet. These bullet points, although correct, may not bear much useful information (e.g., \textit{``Operating system: Android''}).

\begin{table}[H]
\caption{Performance of \tool{} with open-sourced LLMs.}
\label{table:llms}
\resizebox{\linewidth}{!}{
\begin{tabular}{|l|rrrr|}
\hline
                                  & \multicolumn{1}{c}{\textbf{Correctness}} & \multicolumn{1}{c}{\textbf{Uniqueness}} & \multicolumn{1}{c}{\textbf{Overlap}} & \multicolumn{1}{c|}{\textbf{\# of Snippets}} \\ \hline
\textbf{with Llama 3}             & 91.4\%                                   & 77.6\%                                  & 64.7\%                               & 22                                           \\
\textbf{with Llama 2}             & 83\%                                   & 80.3\%                                  & 70.4\%                               & 17.1                                         \\
\textbf{with Mistral}             & 84.9\%                                     & 81.6\%                                  & 65.9\%                               & 21.8                                         \\ \hline
\textbf{\tool{}} & 96\%                                     & 87.6\%                                  & 66.7\%                               & 17.2                                         \\ \hline
\end{tabular}}
\end{table}

\icse{We further examined the API documentation outputs generated by different LLMs to analyze their behavioral similarities and differences.}
\icse{In terms of \textbf{\textit{similarities}}, all models were able to produce well-structured API documentation with seven sections. This suggests that all models can understand the knowledge types explained in our prompt.}
\icse{In terms of \textbf{\textit{differences}}, the models differ in the use of code examples and consistency of formatting. Specifically, GPT-4o and Mistral-7B-v0.3 both included code examples in their outputs (3 and 2 examples respectively across 10 API documents), whereas Llama 2 and Llama 3 did not include any. For consistency, all models except Mistral adhered to a uniform heading format within each section (e.g., using \# or consistent markdown-like markers), while Mistral's output exhibited erratic section header styling. This irregularity may be attributed to Mistral’s smaller model size, leading to reduced consistency in generated documents.}

\subsection{RQ7: Temperature Experiments}
\tool{} by default uses a temperature of 0.8 for GPT-4o to extract, validate, and summarize API knowledge.
Because temperature is a key factor for LLM performance, we did a small-scale analysis on 10 randomly sampled APIs to investigate the \tool{}'s sensitivity to temperature. We used \tool{} to generate documents for the 10 APIs under two other temperature settings: 0 and 0.5. For each API, we ran AutoDoc twice to generate two documents since LLMs may generate different outputs each time being prompted. Thus, 10 APIs × 2 temperatures × 2 trials = 40 documents were generated and labeled by the same two labelers following the same procedure described in Section~\ref{sec:metrics}. Our labeling results (Table~\ref{table:temperature}) show that API documents generated by AutoDoc remain stable across different trials and different temperatures.

\begin{table}[H]

\caption{Temperature experiment results.}
\label{table:temperature}
\begin{center}
\resizebox{0.48\textwidth}{!}{
\begin{tabular}{|l|rrrr|}
\hline
                       & \multicolumn{1}{c}{\textbf{Correctness}} & \multicolumn{1}{c}{\textbf{Uniqueness}} & \multicolumn{1}{c}{\textbf{Overlap}} & \multicolumn{1}{c|}{\textbf{\icse{\#} of Snippets}} \\ \hline
\textbf{T=0/Trial 1}   & 95.3\%                             & 95\%                           & 65.2\%                           & 19                                    \\ 
\textbf{T=0/Trial 2}   & 93.9\%                             & 96.4\%                           & 66.3\%                           & 18                                    \\ 
\textbf{T=0.5/Trial 1} & 95.7\%                           & 95.5\%                           & 62.2\%                           & 21                                    \\ 
\textbf{T=0.5/Trial 2} & 94.5\%                           & 94.2\%                           & 62.9\%                           & 20                                    \\ \hline
\end{tabular}}\end{center}
\end{table}

\section{User Study}
\subsection{User Study Design}
In addition to the quantitative experiments, we conducted a within-subjects user study to evaluate the quality of API documents generated by \tool{}. We recruited {12} students through the CS department mailing list from a local university. Participants had an average of {5.2} years of programming experience.
We randomly sampled 12 APIs from our benchmark.
In each user study, we selected {4} out of the {12} APIs and asked the participants to review the API documents generated by \tool{} and the SISE and GPT-4o (ZSP). We counterbalanced the API document assignment so that \icse{each API document} was evaluated by four different participants.  

For each API, participants were provided with the name of the API, a link to the official API documents, and documents generated by \tool{}, GPT-4o (ZSP), and SISE. The participants were then asked to read all three API documents and evaluate the quality of these summaries by answering the following five multiple-choice questions. The five dimensions were borrowed from~\cite{kou2023automated,7886920}. {For each question, the participants selected one of the API documents generated by SISE, GPT-4o (ZSP), and \tool{}.}

\begin{enumerate}[leftmargin=\parindent]
    \item \textit{Which API document provides the most helpful information?}
    \item \textit{Which API document provides the best overview of the API?}
    \item \textit{Which API document is most comprehensive?}
    \item \textit{Which API document is the most concise without being incomplete?}
    \item \textit{Which API document do you prefer to see in practice?}
\end{enumerate}

 {To mitigate bias, the order of API documents was randomized, and we also did not reveal which model generated \bonan{which of the three API documents}.} 
{At the end of the user study, participants answered two open-ended questions:}

\begin{enumerate}[leftmargin=\parindent]
    \item \textit{Do you wish to see API documents generated from SO posts when learning new APIs?}
    \item \textit{What are ideal API documents like?}
\end{enumerate}

\begin{figure*}[htbp]
    \centering
    \includegraphics[width=0.9\textwidth]{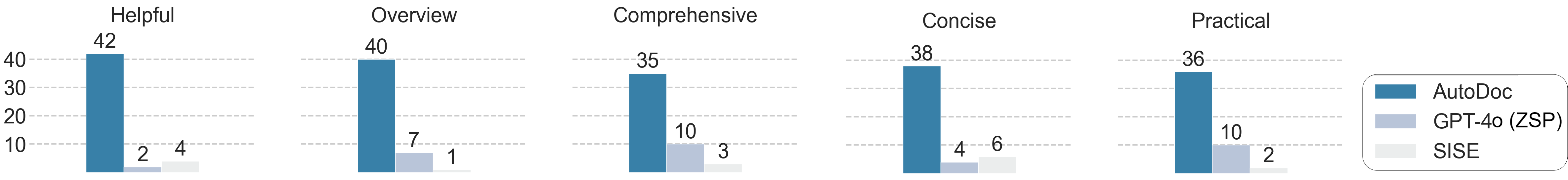}
    \caption{Participants' choices over API documents generated by different methods in five aspects}
    \label{fig:bar_plot}
\end{figure*}

\subsection{User Study Results}

\bonan{Figure~\ref{fig:bar_plot} shows the choice of participants over the API documents generated by \tool{}, GPT-4o (ZSP), and SISE in five aspects. 
On average, 80\% of participants preferred API documents generated by \tool{} over GPT-4o (ZSP) and SISE in all five dimensions. The participants' preference of \tool{} over GPT-4o (ZSP) and SISE is particularly significant in helpfulness (88\%) and overview (83\%). Specifically, P2 said: ``\emph{I particularly enjoy how well-sectioned the API document generated by one of the three methods is (\tool{}).}'' 
P9 said, ``{\em Compared to other methods, this method (\tool{}) generates fewer knowledge snippets that are understandable and not too overbearing.}'' This aligns with our hypothesis since SISE extracts insightful sentences that are sometimes hard to understand without seeing the post, and GPT-4o (ZSP) provides too general and shallow knowledge.

\bonan{In the final survey, 7 out of 12 participants confirmed they would like to see API documents generated from SO posts when working with APIs. For example, P1 said: ``\emph{API documents generated from SO posts contain insights that I have never seen in the official document.}'' {On the other hand, \textit{being well-sectioned} is mentioned by four participants as one of the most important features an ideal API document should possess. For example, P9 said, ``{\em Ideal API documents should be well-structured so I can find what I want easily}''. Meanwhile, P7 said, ``{\em Having well-defined section headers as navigation cues is one of the most important features an ideal API document should possess.}'' Participants also wanted functionalities that \tool{} do not currently possess. For example, two participants said they would like to see the original post from which a knowledge snippet was extracted to have more faith in the generated API document. However, \tool{} cannot provide the post from which a knowledge snippet was extracted because the knowledge summarization component erased all traceability information.

\section{Discussion}
\label{sec: discussion}
In this work, we demonstrate the effectiveness of leveraging insights from Stack Overflow (SO) to generate high-quality API documents. \tool{} can effectively extract API knowledge that is not covered in the official API documents from SO posts. Our evaluation shows that 34.4\% knowledge snippets in API documents generated by \tool{} are not covered in the official API documents. This finding suggests that \tool{} can help developers to learn new APIs with poorly maintained official API documents without having to navigate lengthy SO posts themselves.

\bonan{Furthermore, our evaluation shows the success of \tool{} relies heavily on the knowledge validation component. Hallucination in LLMs' output remains an unresolved concern~\cite{rawte2023survey, zhang2023siren, huang2023survey}. Programmers may not trust LLM-generated API documents because they cannot distinguish the hallucinated content. Our evaluation demonstrates the feasibility of detecting hallucinated content in LLM-generated API documents by leveraging the LLM's self-correction ability. 
Our experiments showed that the knowledge validation component can remove {3.7\%} incorrect knowledge snippets in the generated documents. This result helps to boost user trust in \tool{} and API documents generated by LLMs in general.}

Finally, our experiment shows that only applying a general passage retrieval model, such as the DPR model, is insufficient. When being fine-tuned on the SO data dump, the retrieval accuracy of the DPR model is increased by 13\% on average. This may carry a more general implication---as we reuse models from NLP or ML on SE tasks, we should consider fine-tuning them to account for the unique characteristics of SE tasks. 
For example, SO posts contain unique vocabulary, including many technical terms. Therefore, training the DPR model to understand these terms via fine-tuning can help it better identify posts relevant to an API-related query.


{\bf\em Limitations and Future Work.} First, despite the knowledge validation component, \tool{} still generated API documents with incorrect information, especially for less popular APIs. In practice, it is hard for a developer to distinguish incorrect information from the rest of the document. Therefore, the developers may lose faith in \tool{} and resort to official API documents instead. As a future work, we plan to develop a component that provides traceability to each knowledge snippet (e.g., by providing a link to the original post) to boost user trust. For now, \tool{} does not possess such functionality because, during the summarization process, the LLMs would merge similar knowledge and paraphrase the extracted knowledge. Secondly, many knowledge snippets in API documents generated by \tool{} can be found in the official documents. In the user study, two participants complained about the overlap between the official documents and documents generated by \tool{}. For example, P6 said, \textit{``I think most of the knowledge can be found in official documents. Why don't I just read the official documents?''}. In the future, we would like to develop a knowledge comparison component that automatically filters or marks the shared knowledge between documents by \tool{} and the official documents so developers can only focus on the additional knowledge provided by the auto-generated documents.

\section{Threats to Validity}

In terms of \textbf{internal validity}, the evaluation process involves manual labeling, which can introduce bias in the benchmark. {To mitigate this threat, two labelers iteratively established a set of labeling standards with two labeling rounds. At the end of the second labeling round, they achieved high Cohen's Kappa scores for all three metrics. {Furthermore, although we tried to mitigate data leakage by removing SO posts that mention the 48 APIs, given the size of the dataset--—144K question posts and 175K answer posts, it is impractical to manually verify each post for relevance.}

In terms of \textbf{external validity}, \tool{} was evaluated on 48 APIs with different popularity levels. Therefore, we cannot guarantee that the effectiveness of \tool{} can be generalized to any APIs. In particular, \tool{} may not generalize well for newer or less popular APIs, since these APIs may not be frequently discussed on Stack Overflow. As shown in Table~\ref{table:rq3}, the correctness and uniqueness of knowledge snippets for unpopular APIs drop by 8\% and 7\% and the overlap with official API documentation increases by 12\%, compared to popular APIs. 


In terms of \textbf{construct validity}, the DPR model and GPT-4o can make mistakes. The DPR model may retrieve irrelevant posts. To mitigate this threat, for DPR, we used GPT-4o to check the relevance of posts retrieved by the DPR model. Specifically, we allowed GPT-4o to output nothing in case it cannot extract any API knowledge from the given post. For GPT-4o, we implemented a knowledge \icse{validation} component to detect hallucinations in its output. 

Our user study design also has several limitations. First, our user study may not fully reflect how users would utilize the knowledge extracted by \tool{}, since we only asked participants to choose between several API documents instead of solving real-world problems. \icse{Second, participants were asked to select one preferred document, though sometimes they may not have a strong preference. To mitigate this issue, we randomized the order of the API document options presented to each participant. In this way, even if participants perceived no clear differences and chose randomly, each option would still have the same probability of being selected.}

\section{Related Work}
\subsection{Automated API Documentation}
Several approaches have been proposed for automated API documentation. SISE~\cite{treude2016augmenting} is a framework that augments official API documents with insightful sentences from SO posts. Unlike \tool{}, SISE relies on pattern-based methods that lack flexibility and generate coarse-grained
 API document by extracting original sentences without paraphrasing.
Uddin et al.~\cite{uddin2021automatic} also proposed to automatically produce API documentation from API code examples and reviews from SO posts. Their generated API documents consist of a code example, a textual description, and reviews with positive and negative sentiments.
Stylos et al.~\cite{stylos2009improving} introduced Jadeite, which displays commonly used classes more
prominently and automatically identifies the most common
ways to construct an instance of any given class.
Dekel and Herbsleb~\cite{dekel2009improving} improves API documentation
by decorating method invocations whose targets have associated usage directives, such as rules or caveats. Following a similar method,
Chen and Zhang also proposed to incorporate insights from
crowdsourced FAQs into API documentation~\cite{chen2014asked}. They connected API documents and
informal documentation by capturing the developers’
Web browsing behaviors. 




\subsection{Knowledge Extraction from SE Documents}
There are several approaches that aim to extract API knowledge from API documents and SO discussions~\cite{liu2019generating, li2018improving, wang2019extracting, treude2016augmenting, yin2021api, liu2021api, xu2016predicting}.
Some relied on rule-based pattern matching~\cite{li2018improving, liu2021api}. For example, 
Li et al.~\cite{li2018improving} developed a set of linguistic patterns to extract 10 types of API usage caveats from SO posts. Recently,  neural networks improved the flexibility of some knowledge extraction methods in terms of processing SE documents. 
For example, Liu et al.~\cite{liu2019generating} trained a feed-forward neural network to classify descriptive sentences of APIs from the API documentation.
With a similar idea, DeepTip~\cite{wang2019extracting} uses a CNN model to extract sentence-level API usage tips from SO posts with a trained CNN model.
Compared with previous work, our approach performs knowledge extraction in a finer-grained way because of the pre-trained language models' ability to understand the deep semantics of SO posts. Furthermore, previous neural approaches need to acquire a large labeled dataset first and train a model from scratch. However, our approach makes use of pre-trained models and thus requires no labeled data.

\section{Conclusion}
This paper presents a new API documentation approach called \tool{}. The key insight is to leverage the abundant API knowledge from Stack Overflow to augment the official API documents. Specifically, {\tool} uses a fine-tuned DPR model to identify relevant posts that contain API knowledge from the SO data dump. Then, \tool{} uses GPT-4o to extract API knowledge and perform knowledge validation and summarization to generate a complete API document with seven types of API knowledge.
Evaluation demonstrates that \tool{} can generate API documents that are up to 77.7\% more accurate and 9.5\% more unique than baselines.

\section{Data Availability}
Our code and data have been made available in our GitHub repository~\cite{icse2026_data}.

\begin{acks}
We thank all the anonymous reviewers for their valuable feedback. This work is in part supported by NSF grant ITE-2333736.
\end{acks}

\bibliographystyle{ACM-Reference-Format}
\bibliography{reference}

\end{document}